# Incoherent phonon population and exciton-exciton annihilation in monolayer WS$_2$ revealed by time-resolved spontaneous Raman scattering

Shuangping Han,[1, 3] Christoph Boguschewski,[2] Yan Gao,*[1, 3] Liantuan Xiao,[1, 3] Jingyi Zhu[2] and Paul H. M. van Loosdrecht*[2]

[1] State Key Laboratory of Quantum Optics and Quantum Optics Devices, Institute of Laser Spectroscopy, Shanxi University, Taiyuan, Shanxi 030006, China.

[2] Physics institute 2, University of Cologne, 50937, Germany.

[3] Collaborative Innovation Center of Extreme Optics, Shanxi University, Taiyuan, Shanxi 030006, China.

E-mail: ago@sxu.edu.cn

pvl@ph2.uni-koeln.de

**Abstract**

Atomically thin layer transition metal dichalcogenides have been intensively investigated for their rich optical properties and potential applications in nano-electronics. In this work, we study the incoherent optical phonon and exciton population dynamics in monolayer WS$_2$ by time-resolved spontaneous Raman scattering spectroscopy. Upon excitation of the exciton transition, both the Stokes and anti-Stokes optical phonon scattering strength exhibit a large reduction. Based on the detailed balance, the optical phonon population is retrieved, which shows an instant build-up and a relaxation lifetime of ~4 ps at an exciton density ~$10^{12}$ cm$^{-2}$. The corresponding optical phonon temperature rises by 25 K, eventually, after some 10's of picoseconds, leading to a lattice heating by only ~3 K. The exciton relaxation dynamics extracted from the transient vibrational Raman response shows a strong excitation density dependence, signaling an important bi-molecular contribution to the decay. The exciton relaxation rate is found to be ~ (70 ps)$^{-1}$ and exciton-exciton annihilation rate ~0.1 cm$^2$s$^{-1}$. These results provide valuable insight into the thermal dynamics after optical excitation and enhance the understanding of the fundamental exciton dynamics in two-dimensional transition metal materials.

# 1. Introduction

While the rise of graphene materials [1] and its research are still booming, various other atomically thin materials have emerged, sparking a novel and fascinating field of research [2, 3]. Among them, monolayer transition metal dichalcogenides (TMDCs) play an important role, arising from the intriguing physical properties rooted in their electronically gapped nature, reduced dimensionality, and lack of inversion and time reversal symmetry. TMDCs are regarded by many as a new generation of functional materials with a strong potential for applications in optoelectronic devices [4-9].

One of the significant differences between monolayer TMDCs and graphene is that the former shows an intrinsic band gap, with transition energies ranging from the visible into the near infrared. Rather than free carrier excitations, the lowest energy electronic excitations are strongly bound electron-hole pairs, i.e. excitons with large binding energies up to a few hundred meV [10-13]. These excitonic transitions provide an ideal platform for the study of exciton properties and the related optically induced dynamics in two dimensional semiconductors. This is not only of strong interest from a fundamental materials science point of view, but is also pivotal in eventually realizing the application potential of TMDCs. Not surprisingly, the properties induced by optical excitation and the associated relaxation dynamics in TMDCs have been extensively studied in the recent past using a variety of ultrafast spectroscopic techniques [14-25]. These studies revealed a number of interesting properties, including a giant optically induced band gap renormalization [19, 20, 25] leading to a pronounced spectral dynamics, a strong valley-selective optical Stark effect [26-28], which is of interest in view of optical control in possible valleytronics applications, and a strong material and excitation dependence of the exciton relaxation dynamics showing timescales ranging from tens to hundreds of picoseconds originating from both first and second order relaxation processes [16, 18, 21, 22, 24].

Whereas previous ultrafast spectroscopic studies on monolayer TMDCs mainly focused on the electronic excitation aspects, the lattice and thus the phonon population dynamics, which play a crucial role in the energy dissipation after optical excitation, have not yet been investigated in a direct manner [29]. Time-resolved spontaneous Raman spectroscopy (TRSRS) can provide a direct access to study both the incoherent phonon relaxation as well as the electron-phonon coupling dynamics [30-34]. Recently we have demonstrated that not only the optical phonons but also the exciton population dynamics can be detected by monitoring the optical induced differential Stokes and anti-Stokes Raman signals in one-dimensional graphene nanoribbons [35]. Especially, under a strongly resonant Raman probing condition, the exciton dynamics can be directly obtained from the differential Stokes signal due to its relatively weak dependence on the induced changes in the phonon population. Here we report on an investigation of the phonon creation and relaxation dynamics as well as the exciton relaxation and annihilation mechanisms in monolayer $WS_2$ using TRSRS. The observed fast optical phonon population and depopulation dynamics show an efficient electron-lattice and optical-acoustical phonon coupling leading to fast initial energy dissipation after optical excitation. The observed exciton relaxation time depends strongly on density of optically excited excitons, demonstrating that exciton-exciton annihilation contributes strongly to the exciton relaxation dynamics in monolayer $WS_2$.

## 2. Experimental methods

Monolayer $WS_2$ samples were produced through chemical vapor deposition (CVD) [36, 37] on an oxide coated (280 nm) silicon substrate. The sample was characterized using atomic force microscopy (AFM), steady state photoluminescence and standard Raman spectroscopy, demonstrating the high quality of our sample. The standard Raman and photoluminescence spectra were recorded using a micro-Raman setup equipped with a triple stage spectrometer (Spectroscopy & Imaging GmbH) and $LN_2$ cooled CCD detector (PyLoN 100; Princeton Instruments). A picosecond laser (515 nm, ~ 2 ps) was used for the standard Raman and luminescence measurements. The 515 nm laser pulse was spectrally cleaned and narrowed (full width a half maximum (FWHM) ~10 cm$^{-1}$) using a home build pulse shaper. The laser pulses were focused on the sample using microscope objective (20×, NA=0.4). The Raman or photoluminescence signals of $WS_2$ were collected in a backscattering geometry.

Details of the TRSRS technique used here and related data analysis methods have been described elsewhere [38]. Briefly, an integrated ultrafast laser system (Light Conversion PHAROS) with two outputs of the fundamental pulses (300 fs and 150 ps, @1030 nm) pump two optical parametric amplifiers (Light Conversion), one to generate laser pulses for selective excitation (~300 fs), and one to generate a narrow-bandwidth laser pulse for Raman probing(~2 ps). For excitation of the $WS_2$ sample, the pump pulse wavelength was centered at ~ 620 nm (2.0 eV), corresponding to the direct A exciton resonance. The Raman probe pulse was set at 515 nm (2.4 eV), in resonance with the B excitonic transition [12]. This configuration of pump and probe yields a strong resonant Raman probing signal and a direct excitation of excitons with only a small amount of excess energy. In order to minimize the background scattering induced by the relatively strong pump pulse, a crossed linear polarization of pump and probe was used, while detecting the Raman signal polarized parallel to the probe pulse.

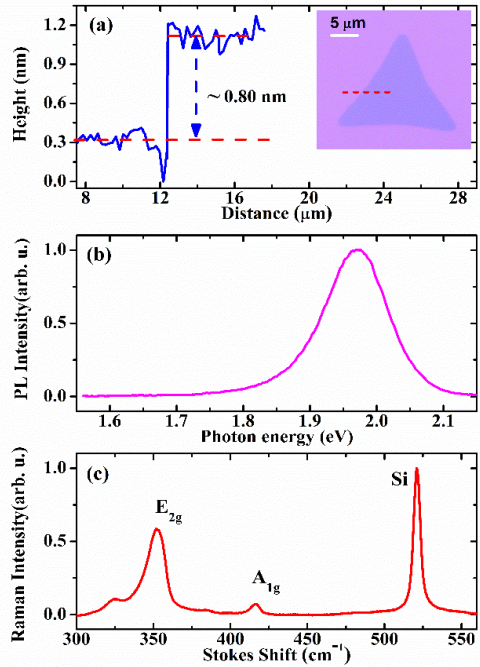

Fig. 1 Steady state spectra characterization of the CVD grown monolayer $WS_2$ sample on silica/silicon substrate. (a) Atomic force microscope (AFM) measurement the thickness of monolayer $WS_2$, inset show the optical image and AFM scanning range (dashed red line). (b) Photoluminescence. (c) Steady Raman scattering.

## 3. Results and discussion

The inset of Fig. 1(a) shows an optical image of the triangularly shaped crystalline monolayer $WS_2$ sample. The main panel 1(a) presents an AFM line scan along the dashed line in the inset image, showing the for a monolayer expected thickness of roughly 0.8 nm [39]. The monolayer quality of the sample was further confirmed by the steady state photoluminescence (PL) and Raman spectra presented in Fig. 1 (b) and (c), respectively. The PL spectrum has a peak located around 1.97 eV, corresponding to the excitonic A transition of crystalline single layer $WS_2$ [39, 40]. The slight broadening of the PL spectrum and the asymmetric peak shape with weaker tails extending towards the lower energy side is due to structural defects inducing different charged exciton contributions [41-43]. The Raman spectrum shows a strong response at 350 cm$^{-1}$, corresponding to an E$_2$g in-plane optical phonon mode and a weaker mode at 415 cm$^{-1}$ originating from either the out-plane vibration or from defect modes [44-46]. The peak located at 520 cm$^{-1}$ is due to the underlying silica/silicon substrate. The peak positions and the relative ratio of peak intensities are in good agreement with previously reported Raman spectra of monolayer WS$_2$ on silica/silicon substrate [47, 48]. In the time-resolved measurements, we focus on the dynamic changes induced on the strongest in-plane mode (E$_2$g) at around 350 cm$^{-1}$.

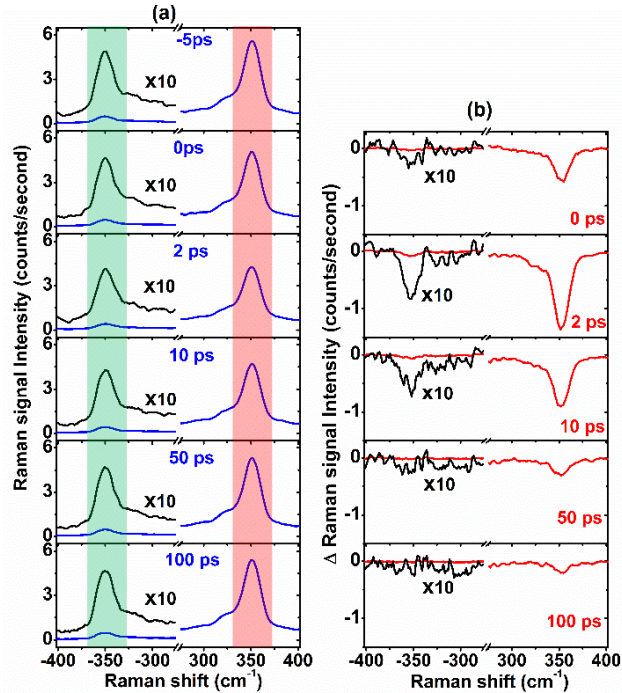

Fig. 2 Time-resolved spontaneous Raman scattering spectra of the monolayer WS$_2$ on silica/silicon recorded on both Stokes and anti-Stokes sides simultaneously. (a) Raman scattering intensity spectra at different delay times after optical pump at 2.0 eV. (b) Pump induced difference spectra obtained by subtraction the spectrum at -5 ps from each spectrum in (a) at different corresponding delay times.

Time-resolved spontaneous Raman scattering spectra of monolayer WS$_2$ around the 350 cm-1 E$_2$g phonon response are presented in Fig. 2. Fig. 2(a) shows the transient Stokes and anti-Stokes spectra for various pump-probe delay times using an initial photo-excited exciton density of ~ 2.26 × 10$^{12}$ cm$^{-2}$, whereas Fig.2(b) shows the derived difference spectra obtained by subtracting the response at -5 ps (top panel Fig. 2(a)). From these spectra it is clear that the optical pump induces a significant reduction of the scattering efficiency, without any spectral dynamics of broadening or shifting, which recovers on a time scale of tens of ps. As we have discussed previously for the resonant Raman scattering case [35] the induced transient changes may originate both from changes in the resonant enhancement due to electronic population effects (typically leading to a reduction of the response) as well as from changes in the vibrational population (typically leading to an increase of the response). Normally semiconductors show an induced increase of the anti-Stokes Raman signals [30, 31, 33-35, 38, 49, 50] assigned to a dominated contribution of optical phonon populations, and an increase or decrease of the Stokes scattering depending on resonance conditions. For monolayer $WS_2$, however, this is not the case. Here both Stokes and anti-Stokes scattering show a transient decrease, suggesting a dominating contribution from transient changes in the electronic resonance enhancement overshadowing the phonon population contribution on both sides.

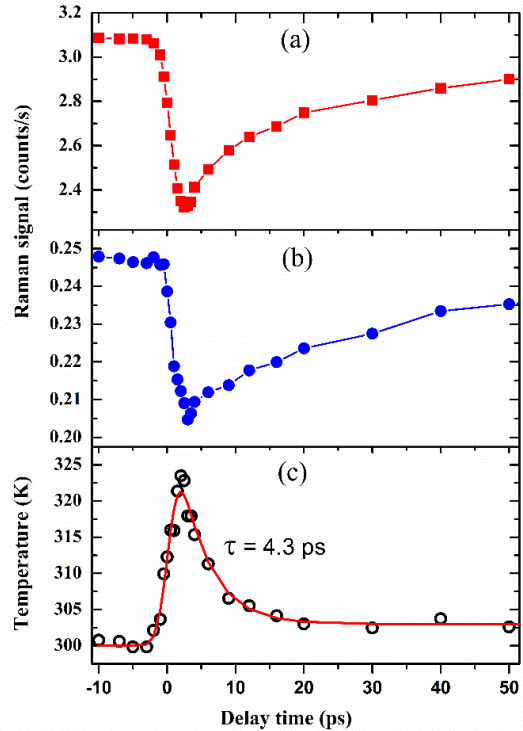

Fig. 3 Decay dynamics of the optical phonon peak at around 350 cm$^{-1}$ at the exciton density around $2.26 \times 10^{12}$ cm$^{-2}$. (a) Stokes side dynamics integrated in the spectral region from 338 to 362 cm$^{-1}$. (b) Anti-Stokes side dynamics integrated in the spectral region from -362 to -338 cm$^{-1}$. (c) Calculated temperature of the optical phonon according to Bose-Einstein statistics.

To obtain a better view on the detailed relaxation dynamics of the optical phonon scattering, the $E_{2g}$ phonon peak at ~ 350 cm$^{-1}$ was spectrally integrated (green and red bar indicated in Fig. 2(a)) for different delay times. The obtained decay dynamics are presented in Fig. 3 for the Stokes (a) and anti-Stokes (b) responses. Though both anti-Stokes and Stokes sides show a transient reduction with a recovery time of some 10s of ps, the detailed transient response is substantially different, in particular at early times. The origin of this difference is the (positive) contribution of the phonon population to the response, in particular for the anti-Stokes signal. In order to retrieve the optical phonon population dynamics, we assume that shortly (<1ps) after optical excitation the fluctuation-dissipation theorem holds and the ratio between Stokes and anti-Stokes scattering is given by

$$I_S /I_{AS} = \eta * \exp (h\Omega/k_B T), \qquad (1)$$

in which h is the plank constant, $\Omega$ is the optical phonon frequency, $k_B$ is the Boltzmann constant and T is the phonon temperature. The prefactor $\eta$ summarizes differences in for instance resonance enhancement [51, 52] and optical material properties [53, 54], and can be determined from an experiment at a known temperature. To determine $\eta$, we further make the assumption that before time zero, the phonon temperature is close to the environment (~ 300 K). This assumption is reasonable in view of the low Raman probe laser pulse fluence, and the observation that there are no significant differences between pre-zero delay time intensities measured with and without the presence of the pump laser beam.

The transient phonon temperature determined using Eq. (1) presented in Fig. 3(c) shows an instantaneous (within our time resolution) increase by about 25 K, corresponding to a transient increase of the optical phonon population. The increased phonon population relaxes with a decay time of 4.3 ps to a long lived quasi equilibrium state with a temperature of about 3 K above base temperature, which further relaxes to the base temperature outside our detection time window but within the time between two pump pulses (10 μs).

The observation of the fast (<2ps) rise in phonon temperature after excitation suggests that optical phonon populations are created through a fast exciton-phonon scattering process, This can be either thorough release of the excess energy of the optically excited exciton states, or through transitions from the optically excited bright exciton states into optically inactive dark exciton states. In monolayer $WS_2$ the dark exciton state is formed by the electrons in the conduction band and holes in the valence band with opposite spins, and the energy of this dark exciton state is slightly lower (30-50 meV) than the bright exciton state [55-59]. Since relaxation from the bright to the dark exciton state requires a spin flip of the electrons, this process is expected to be substantially slower than the exciton cooling process but cannot be excluded on the basis of the current experiments. The relaxation of the optical phonon population (~ 4.3 ps) can be ascribed to optical-acoustic phonon scattering usually occurring on a timescale of a few ps in semiconductors [30, 31, 33-35, 38, 49].

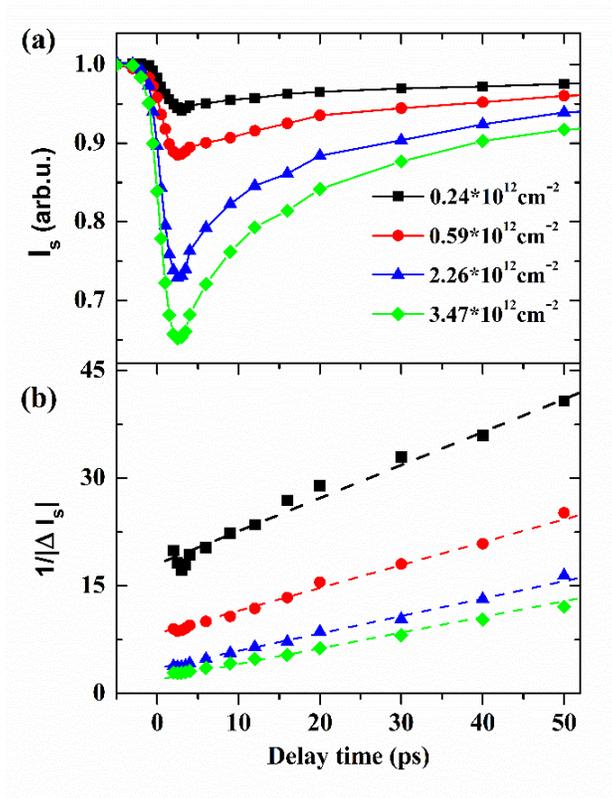

Fig. 4 Intensity dependence of the dynamics observed on Stokes side. (a) Decay dynamics of the optical phonon peak at around 350 cm$^{-1}$ at different exciton density. (b) Inversion of the decay dynamics (dots) from (a) and the global fitted ones (dashed lines) with rate equation including both the first and second order exciton annihilation reaction (details see text in the paper). The global fitting extracted rate constants of k1 ~ (67 ps)$^{-1}$ and k$_2$ ~ 0.104 cm$^2$s$^{-1}$.

Apart from information on the phonon population dynamics and lattice temperature, one can straight forwardly derive the exciton population dynamics from the resonant TRSRS experiment. This is done by analyzing the time resolved optical phonon Stokes scattering signal, which is hardly influenced by the minor changes in total phonon population induced by the pump pulse.[35,38] The observed transient changes in the Stokes response can be ascribed to changes in the resonance enhancement due to ground state bleaching/excited state filling by the pump pulse, i.e. due to a transient reduction of the optical transition probability. The 10s of ps recovery time of the phonon signals is therefore assigned to relaxation of the excited excitonic states. In order to get a better insight into the exciton relaxation dynamics we performed a set of experiments for varying initial exciton densities. Fig. 4a shows the integrated $E_{2g}$ mode Stokes side signal for different excitation densities. The data show a clear speed up of the recovery dynamics upon increasing excitation density, indicating that many body annihilation processes play a role in the decay dynamics.

To analyse the observed phenomena, we model the dynamics by including both a first order free exciton decay process and a second order relaxation process, i.e., exciton-exciton annihilation. The population decay can then be expressed as:

$$dN(t)/dt = -k_1 N(t) - k_2 N^2(t), \qquad (2)$$

where $N(t)$ is the time dependent exciton population density, and $k_1$ ($k_2$) is the first (second) order rate constant. Analytically solving (2) gives

$$1/N(t) = [1/N(0) + k_2/k_1]\exp(k_1 t) - k_2/k_1, \qquad (3)$$

in which $N(0)$ is the initially excited exciton density. Since the effective lifetime from the first order reaction is usually much longer than that from the second order one, to simplify the description, in short time range, we expand $\exp(k_1 t)$ to $1+k_1 t$ and (3) becomes:

$$1/N(t) = 1/N(0) + [k_1/N(0) + k_2]t \qquad (4)$$

Expression (4) gives a very compact and intuitive description for the observed signals: it indicates that the inverse of the differential Raman signals should be simply linearly proportional to the delay time t, while the slope is excitation density dependent. Indeed, as shown in Figure 4(b), this is exactly the case here. Global fitting (dashed lines) of the data (symbols) using Expression (4) yields satisfactory agreement and a first order rate constant $k_1 = (67\ \text{ps})^{-1}$ and second order $k_2 = 0.104$ $cm^2 s^{-1}$. These values are comparable to those reported for other measurements on CVD grown monolayer $WS_2$ samples [21, 22], confirming the importance of an efficient exciton-exciton annihilation process in monolayer $WS_2$. The slight difference of the rate constants observed here and those in previous experiments on CVD grown samples are most likely due to variations in defect densities. We note that the values reported for CVD grown samples differ substantially from those measured on exfoliated samples. In exfoliated monolayer $WS_2$, the first order exciton decay rate constant was measured to be around $(806\ \text{ps})^{-1}$, one order slower, while the exciton-exciton annihilation rate $k_2$ was around $0.41\ cm^2 s^{-1}$, 4 times larger [18]. The typical defect density (~ $3\times10^{13}$ $cm^{-2}$) [60] in a CVD grown sample is usually much higher than that in exfoliated samples (~ $2\times10^9$ $cm^{-2}$) [61], This difference strongly influences the diffusion of excitons [62] and thus affects both the first and second order decay processes, leading to a larger $k_1$ due to defect assisted recombination [63, 64] and a smaller $k_2$ due to slower diffusion in CVD grown $WS_2$. These results suggest that the quantum efficiency of light emission and thermal effects caused by strong excitation conditions may be optimized for optical device applications by manipulating the defects density in the fabrication process of TMDCs.

## 4. Conclusion

In conclusion we have investigated the relaxation dynamics of both the phonons and excitons after optical excitation in a CVD grown monolayer $WS_2$ using TRSRS spectroscopy. The population the $E_2g$ optical phonon shows a very fast (<2 ps) increase due to efficient exciton-phonon coupling and decays with time constant of around 4 ps through optical-acoustic phonon scattering. The exciton relaxation and annihilation dynamics, as observed through the time dependent Stokes optical phonon signals show the presence of both first and second order decay processes. The first order and the second order exciton decay rates obtained from the experiments are consistent with those obtained by other methods for CVD grown samples, but differ substantially from those for exfoliated samples due to the higher defect density. The present results provide a direct insight into both the excitonic and vibrational energy dissipation properties of TMDC materials which has relevance to potential electro-optical TMDC applications, in particular under strong optical or electronic excitation conditions.

**Acknowledgements**

The authors acknowledge financial support funded by the Deutsche Forschungsgemeinschaft (DFG, German Research Foundation) - Project number 277146847 - CRC 1238 and the support from the National Key Research and Development Program of China [Grant No. 2017YFA0304203]; National Natural Science Foundation of China (NSFC) [Grant Nos. 61605104, 61527824, 11434007]; PCSIRT [Grant No. IRT_17R70]; 111 project [Grant No. D18001] and 1331KSC; the Applied Basic Research Program of Shanxi Province (No. 2016021017).